\documentclass[%
 reprint, 
superscriptaddress,
 amsmath,amssymb,
 aps, 
pra,
]{revtex4-2}

\usepackage{dsfont}
\usepackage[dvipsnames]{xcolor}
\usepackage{tikz}

\usepackage{placeins}
\usepackage[T1]{fontenc}

\usepackage{physics}

\usepackage{centernot}
\usepackage{mathtools}
\usepackage{stmaryrd}

\usepackage{subcaption}
\usepackage{capt-of} 
\usepackage{todonotes}
\usepackage{svg}
\usetikzlibrary{decorations.pathreplacing}
\usetikzlibrary{arrows}
\usetikzlibrary{calc}
\usetikzlibrary{shapes.geometric}
\usetikzlibrary{decorations}
\usetikzlibrary{arrows.meta}
\usetikzlibrary{fadings}
\tikzfading[name=fade out, 
    inner color=transparent!0,
    outer color=transparent!100]
\pgfdeclaredecoration{lightning bolt}{draw}{
\state{draw}[width=\pgfdecoratedpathlength]{
  \pgfpathmoveto{\pgfpointorigin}%
  \pgfpathlineto{\pgfpoint{\pgfdecoratedpathlength*0.6}%
    {-\pgfdecoratedpathlength*.1}}%
  \pgfpathlineto{\pgfpoint{\pgfdecoratedpathlength*0.55}{0pt}}%
  \pgfpathlineto{\pgfpoint{\pgfdecoratedpathlength}{0pt}}%
  \pgfpathlineto{\pgfpoint{\pgfdecoratedpathlength*0.4}%
    {\pgfdecoratedpathlength*.1}}%
  \pgfpathlineto{\pgfpoint{\pgfdecoratedpathlength*0.45}{0pt}}%
  \pgfpathclose%
}%
}

\makeatletter
\newcommand{\xMapsto}[2][]{\ext@arrow 0599{\Mapstofill@}{#1}{#2}}
\def\Mapstofill@{\arrowfill@{\Mapstochar\Relbar}\Relbar\Rightarrow}
\makeatother

\usepackage{siunitx}
\usepackage{graphicx}
\usepackage{dcolumn}
\usepackage{bm}
\usepackage{hyperref}
\usepackage{cleveref}

\newcommand{\amherst}{Manning College of Information and Computer Sciences, University of Massachusetts Amherst, 140 Governors Dr, Amherst, MA 01002, United States}

\newcommand{\ulm}{Institute for Complex Quantum Systems, Ulm University, 89069 Ulm, Germany}
\newcommand{\iqst}{Center for Integrated Quantum Science and Technology (IQST), Ulm-Stuttgart, Germany}
\newcommand{\irvine}{Department of Mathematics, University of California, Irvine, CA, USA}
\newcommand{\siegen}{Naturwissenschaftlich-Technische Fakult{\"a}t, Universit{\"a}t Siegen, Walter-Flex-Stra{\ss}e 3, 57068 Siegen, Germany}

\begin{document}

\title{\textbf{Exact noise characterization of entanglement distribution in star networks} 
}%

\author{Kenneth Goodenough}
\affiliation{\amherst}
\affiliation{\siegen}
\email{Contact author: kdgoodenough@gmail.com}
\author{Xiaonan Chen}
\affiliation{\irvine}
\affiliation{\amherst}
\affiliation{\siegen}
\author{Patrick Emonts}%
\affiliation{\ulm}
\affiliation{\iqst}

\date{\today}
\begin{abstract}
Multipartite entanglement forms the core of many networking applications. 
In the near-term future, it is expected that multipartite distribution will be achieved first through star topologies, making it important to understand the noise incurred during the distribution process.
In such networks, elementary links are created stochastically and successful links must be stored while waiting for the remaining links, causing memory decoherence that depends on the random waiting times. 
We derive analytical expressions for both the average noise and its distribution, when distributing GHZ states under memory dephasing in star networks. 
We study and compare two distribution protocols: the factory and piecemaker protocol. 
Furthermore, we find expressions for the case of a global cut-off (allowing fast optimization of the cut-off without requiring Monte Carlo simulations) and extend the analysis for the factory protocol to depolarizing noise for arbitrary states.
\end{abstract}

\maketitle

\section{Introduction}

Multipartite entanglement is a natural primitive for multi-user tasks in a quantum internet, ranging from conference key agreement~\cite{murta2020quantum} and secret sharing~\cite{markham2008graph, gravier2012quantum} to distributed sensing~\cite{zhang2021distributed} and clock synchronization~\cite{kong2018demonstration, lamas2018secure}. 
In essentially all of these settings, the relevant figure of merit is the quality of the end-to-end entanglement. 
This makes it crucial to understand how protocol design and physical noise processes determine the quality of the final delivered state~\cite{goodenough_noise_2024, shchukin2019waiting, li2021efficient}.

A key feature of near-term networks is that entanglement generation is heralded and stochastic: elementary links succeed at random times, and the resulting entangled pairs must typically be stored until other links have succeeded, such that they can be jointly measured to establish larger states. 
Consequently, the noise affecting the final multipartite state is necessarily probabilistic, and thus has an associated distribution. 

From a practical standpoint, closed-form characterizations of this distribution (such as its mean, moments, or even the full distribution) are valuable for at least two reasons. 
First, they enable rapid parameter sweeps and protocol optimization (e.g., the number of users $n$, link success probability, and memory coherence), without relying on expensive Monte Carlo simulations. 
Second, they provide insights usually not provided by numerics or simulations, such as scaling behavior in the underlying experimental parameters.

As such, it is important to understand the noise in quantum networks as thoroughly as possible. 
Previous works have analytically studied the noise in swap ASAP repeater chains~\cite{kamin_exact_2023, goodenough_noise_2024}, and sequential distribution of $n$ Bell pairs between two nodes~\cite{chehimi2025entanglement}, among others. 
Here we continue this line of investigation by characterizing the noise in star networks \emph{exactly}, meaning we find analytical expressions for the average fidelity for arbitrary $n$, and the distributions analytically for small numbers of users $n$, and numerically for larger numbers of users $n$.

The random variable in our analysis is the accumulated storage time $K$ of each qubit, which in turn gives rise to a noise parameter $\Lambda=\lambda^K$. 
Our goal is to characterize both the average effective noise, $\mathds{E}[\Lambda]$, and the distribution of $K$, from which the distribution of the delivered-state fidelity can be obtained. 

We consider two protocols for distributing entanglement: the factory and piecemaker protocol. 
The factory protocol~\cite{avis2023analysis} attempts elementary link generation until all links incident to a network node have succeeded, after which a GHZ-basis measurement is performed at that network node.
This requires the network node to store the Bell pairs until they have all arrived, incurring a significant amount of decoherence. 
The piecemaker protocol~\cite{prielinger2025piecemaker} reduces the amount of decoherence by exploiting the fact that a GHZ-basis measurement can be split into a number of smaller measurements, which can be performed on subsets of Bell pairs. 
This allows for certain Bell pairs to be measured out before all Bell pairs have arrived, reducing the amount of decoherence experienced. 
We give more details on both these protocols in Section~\ref{sec:protocols}. 
We note that these two protocols far from exhaust the possible protocols in star networks, see e.g.~\cite{sen2025multipartite, cuquet2012growth}.

Our main contributions are as follows. 
First, for homogeneous star networks under dephasing noise, we find closed-form expressions for the average noise of both the factory and piecemaker protocols (section \ref{sec:exp_value}). 
Second, we extend the factory analysis to arbitrary target states under uniform depolarizing noise (appendix \ref{sec:dep_factory}), improving on the results from~\cite{avis2023analysis}. 
Third, we obtain expressions for both protocols for finite global cut-offs, enabling efficient optimization of quantities such as the conference-key rate (section \ref{sec:cut_off_analysis}). We also provide a closed-form expression for the average generation time when using a global cut-off (appendix \ref{app:waiting_times}).
Finally, we show that the average noise, viewed as a probability generating function, determines the full distribution of the accumulated storage time. Such distributions are useful when using so-called `binning' approaches, which can significantly improve conference-key agreement rates (section \ref{sec:distribution}).

We note that the analysis here is for the highly symmetric case of a homogeneous star network.
We show in our follow up paper~\cite{goodenough2026optimization} that one can reframe our results in terms of tensor networks, which allows us to numerically compute and optimize the expectation values for not only the inhomogeneous setting~\cite{memmen2026strategy}, but also for more complex network topologies. Our code is publicly available at Ref.~\cite{github_repeater-network-analytics}.

\section{The protocols}\label{sec:protocols}
\subsection{Factory protocol}
The factory protocol~\cite{avis2023analysis} attempts elementary link generation with the $n$ end users until all elementary links have been generated. 
Afterwards, a GHZ-basis measurement is performed on the qubits held by the central node, effectively distributing a GHZ state to the end users (after they have performed corrections depending on the outcome of the measurements). 
We note that Ref.~\cite{avis2023analysis} studied the average noise in the regime of low noise and high elementary link success probability, which they numerically confirmed to be tight in practice for those regimes. 
We close this gap by providing an exact expression for all parameters. 
Moreover, we find expressions for the average \emph{state}, and we find the \emph{distribution} of the noise, as we show in Section~\ref{sec:distribution}. 
We note that other protocols were also considered in Ref.~\cite{avis2023analysis}, but these are outside the scope of this paper.

\subsection{Piecemaker protocol}
The factory protocol is based on a number of operations that---by themselves---act on at most two qubits. 
Because of this, operations can be performed before all Bell pairs have arrived. 
The piecemaker protocol~\cite{prielinger2025piecemaker} exploits this capability to perform measurements as soon as possible. 
In particular, a number of measurements can be performed before all Bell pairs have arrived, reducing both memory decoherence and freeing up the memories for potential other applications. Which measurements can be performed earlier depends both on the Bell pairs that have arrived, and on the entanglement structure of the stabilizer state to be distributed.

We give now a brief description of the piecemaker protocol for the specific case of a GHZ state, which is particularly simple: the GHZ-basis measurement itself can be naturally decomposed into smaller fusion measurements (see Ref.~\cite{prielinger2025piecemaker} for further details).

First, entanglement generation attempts are performed for every end user. 
After the first Bell pair arrives, the associated qubit is kept in storage until the end. 
Afterwards, every subsequent Bell pair that arrives can be fused with the first qubit through a type 1 fusion measurement. 
After all Bell pairs have arrived, the qubit of the initial Bell pair is measured, and a final correction is performed by the end users.

As shown through simulations in Ref.~\cite{prielinger2025piecemaker}, allowing the measurements to be performed earlier indeed increases the average state quality, where this benefit is in particular the strongest for GHZ states. 
Here we provide the analytical analysis of the piecemaker protocol, albeit only for the case of dephasing noise on GHZ states.

\section{Noise model}\label{sec:noise}
We restrict ourselves here to the setting where the only noise is given by memory decoherence. 
In particular, we do not take into account gate/measurement noise and initial state infidelities. 
While the former is significantly harder to capture analytically, it is straightforward to deal with the latter (see e.g. Ref.~\cite{goodenough_noise_2024}).

Furthermore, we consider only the case of decoherence on the qubits at the central node. 
This is motivated by the fact that for a number of practical protocols, the end users can immediately measure out their qubits after a success~\cite{murta2020quantum}.

We consider both dephasing and depolarizing noise in this work, which are represented by the following respective noise maps (when restricted to single qubits)

\begin{align}
\rho \xmapsto[]{~\lambda~}\left(\frac{1+\lambda}{2}\right)\rho + \left(\frac{1-\lambda}{2}\right)Z\rho Z^\dagger\, \textrm{ and}\\
\rho \xmapsto[]{~\lambda~} \lambda\rho + \left(1-\lambda\right)\mathrm{Tr}\left(\rho\right)\frac{\mathds{I}}{2} \ .
\end{align}

Let us momentarily restrict to dephasing when distributing a GHZ state with the factory protocol.

First, it suffices to apply the average noise accrued at the central node to the state at the end users, where we use the same `teleportation' argument as in Ref.~\cite{avis2023analysis}. 
Second, we make use of the fact that dephasing affects GHZ states in a simple manner; if $\rho_{\pm}$ are the density matrices of $\frac{\ket{0}^{\otimes n } \pm \ket{1}^{\otimes n}}{\sqrt{2}}$, then the resultant state is of the form

\begin{align}
\left(\frac{1+\Lambda}{2}\right)\rho_{+} + \left(\frac{1-\Lambda}{2}\right)\rho_{-} \, \label{eq:avg_state},
\end{align}
and $\Lambda=\lambda^k$, where $k$ is the total number of times the dephasing map with (fixed) $\lambda$ has been applied to the state. In particular, it does not matter on which qubits the dephasing map was applied --- only the total number of rounds for which the qubits were stored matters.

This motivates us to define $\Lambda=\lambda^k$ as the total noise parameter, where in particular the average noise parameter 

\begin{align}
    \mathds{E}\left[\Lambda\right] =& \sum_{k=0}^\infty \mathrm{Pr}\left[K=k\right] \cdot \lambda^k \label{eq:gen_expr1}\\
    =&\sum_{\overline{t}} \textrm{Pr}\left[~\overline{t}~\right] \cdot\Lambda\left(\overline{t}\right)\label{eq:gen_expr2} 
\end{align}
is of interest, where $K$ is the random variable corresponding to the total storage time of the qubits, and $\overline{t}=\left(t_1, t_2, \ldots, t_n\right)$ describes an instance of the protocol. Here the $t_i\in \mathbb{N}_{>0}$ are the times at which each end user $i$ generates entanglement with the central node. From Eq.~\eqref{eq:avg_state} and linearity of the expectation operator, it follows that from $\mathds{E}\left[\Lambda\right]$ the average state and average fidelity (i.e.~$F=\frac{1+\mathds{E}\left[\Lambda\right]}{2}$) can be straightforwardly calculated. 

One can similarly extend this logic to the GHZ variant of the piecemaker protocol, since fusing two GHZ states together with noise parameters $\Lambda_1$ and $\Lambda_2$ yields a GHZ state with noise parameter $\Lambda_1\cdot \Lambda_2$. This follows from teleporting the noise away from the qubits being fused, ensuring that the fusing and noise maps commute.\\

It follows that, in the case of dephasing noise, the noise in both the factory and piecemaker protocol are captured by expressions of the form of Eqs.~\eqref{eq:gen_expr1} and~\eqref{eq:gen_expr2}. We specialize these expressions in the following section, and derive explicit closed-form expressions for them in Appendices~\ref{sec:factory} and~\ref{app:piecemaker_deriv}.\\

Finally, it is possible to extend the analysis for the factory protocol to the case of arbitrary states and uniform depolarizing noise, see Appendix~\ref{sec:dep_factory} for further details. The argument is based on expanding the target state in the Pauli basis (i.e.~all phaseless Pauli strings $P$ of length $n$), and noting that depolarizing noise is a diagonal map $P\mapsto \Lambda_P P$ in this basis. It thus suffices to understand $\mathds{E}\left[\Lambda_P\right]$ for each $P$. Furthermore, since depolarizing noise is symmetric and we consider a homogeneous star network, $\mathds{E}\left[\Lambda_P\right]$ depends only on the weight $w(P)$ of $P$, i.e.~the  number of non-identity entries in $P$. We provide a closed-form expression of  $\mathds{E}\left[\Lambda_P\right]$ as a function of $w(P)$ in Appendix~\ref{sec:dep_factory}, see Eq.~\eqref{eq:factory_avg_dep}.

Note that a similar argument cannot be used to extend depolarizing to the piecemaker protocol, since depolarizing noise cannot be straightforwardly teleported between the qubits anymore, even when restricted to GHZ states. Furthermore, the piecemaker protocol requires two-qubit gates mid-distribution for more complex stabilizer states, significantly complicating the analysis, even when restricting to dephasing only. It is for this reason that we restrict our analytics for the piecemaker protocol to dephasing noise on GHZ states. With this in mind, we will only consider dephasing noise on GHZ states in the following section (unless otherwise noted), to ensure a fair comparison between the two protocols.

\section{Results}
In this section, we apply our tools to understand the noise while distributing entanglement in star networks, for both the factory and piecemaker protocol. 
In the first subsection, we study the expectation value of the noise.
The second subsection is devoted to the effects of cut-offs on the expectation value of the noise, the generation rate, and the rate at which one can perform conference key agreement~\cite{murta2020quantum}. 
We conclude with a subsection discussing the distribution of the noise, and apply this to a `binned' version of conference key agreement.

\subsection{Expectation value of the noise}\label{sec:exp_value}
Here we study the expectation value of the noise parameter. 
To do so, we take the underlying models of our protocols, and express the corresponding expectation values as an infinite sum. 
In Appendices~\ref{sec:factory} and~\ref{app:piecemaker_deriv} we find closed-form expressions of both these infinite sums as sums over $n+1$ terms.

We will start with the piecemaker protocol, as it is conceptually the simplest. 
Note that an instance of the piecemaker protocol is described by a sequence of strictly positive integers $\overline{t} = \left(t_1, t_2, \ldots, t_n\right)$ (and that each such $\overline{t}$ describes an instance), where each $t_i$ is the round in which the $i$'th Bell pair succeeded. 
The piecemaker protocol has only one Bell pair decohering, namely the one that arrived first. 
This Bell pair is stored from the time it is generated, until the last Bell pair gets generated. 
As such, the total waiting time is $\max(\overline{t}) - \min(\overline{t})$, leading to a final noise parameter of $\Lambda=\lambda^{\max(\overline{t}) - \min(\overline{t})}$. 
The probability of observing such an instance is $\textrm{Pr}\left[\max(\overline{t})=a\land \min(\overline{t})=b \right]$, such that

\begin{align}\label{eq:piecemaker_eq}
\mathds{E}\left[\Lambda\right]=\sum_{a\geq b}\lambda^{a-b}~\textrm{Pr}\left[\max(\overline{t})=a\land \min(\overline{t})=b \right] \ . 
\end{align}
We provide a closed-form expression of this sum in Appendix~\ref{app:piecemaker_deriv}.

Let us move on to the factory protocol. 
In this case, each Bell pair $i$ is generated at time $t_i$ and then stored until $\max(\overline{t})$, at which point it gets measured out. 
This leads to a total storage time of $\max(\overline{t})-t_i$ for each Bell pair $i$. 
Since the noise is multiplicative (see Section~\ref{sec:noise}), the noise parameter $\Lambda$ for a given instance $\overline{t}$ is given by 

\begin{align}
    \Lambda=\lambda^{\sum_{t_i \in \overline{t}}\left(\max(\overline{t})-t_i\right)} \ .
\end{align}
The probability of observing an instance $\overline{t}$ is given by

\begin{align}
\textrm{Pr}\left[\overline{t}\right] =& \prod_{t_i\in \overline{t}}p(1-p)^{t_i-1}\\
=&\left(\frac{1-q}{q}\right)^nq^{\sum_{i=1}^nt_i} \ ,
    \end{align}
    where we have used that the generation of $n$ Bell pairs follows the same distribution as $n$ geometric random variables, and where we have set $q \equiv 1-p$. Combining the above two expressions yields
\begin{align}\label{eq:factory_eq}
\mathds{E}\left[\Lambda\right] =& \sum_{\overline{t}} \textrm{Pr}\left[~\overline{t}~\right] \cdot\Lambda\left(\overline{t}\right)\\
=&\left(\frac{1-q}{q}\right)^n\sum_{\overline{t}} q^{\sum_{i=1}^nt_i} \lambda^{\sum_{t_i\in\overline{t}}\left(\max(\overline{t})-t_i\right)}\ . 
\end{align}
We provide a closed-form expression of the above sum in Appendix~\ref{sec:factory}, see Eq.~\eqref{eq:factory_avg}.

\begin{figure}[h]
    \centering
    \includegraphics[width=1\linewidth]{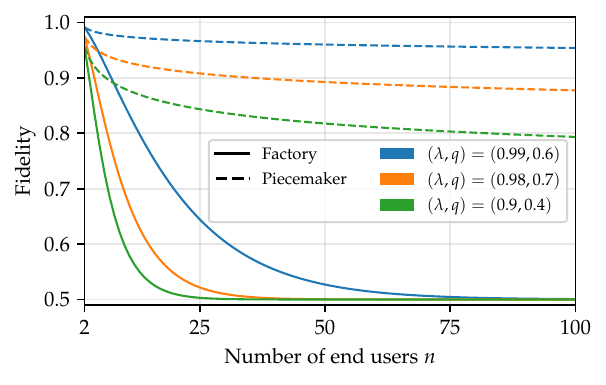}
    \caption{Fidelity as a function of the number of end users $n$, for both the piecemaker and factory protocol. 
    The solid and dashed lines correspond to the factory and piecemaker protocol, respectively. We plot the fidelity for several experimental parameters $(\lambda, q) = (0.99, 0.6), (0.98, 0.7)$ and $(0.9, 0.4)$.}
    \label{fig:fid_vs_n}
\end{figure}

With closed-form expressions for the average noise parameters of the piecemaker and factory protocols in hand, we now compare their performance in Fig.~\ref{fig:fid_vs_n}, where we plot the fidelity as a function of the number of end users and fixed values of $\left(\lambda, q\right)$. We observe that the piecemaker protocol outperforms the factory protocol, consistent with the results from~\cite{prielinger2025piecemaker}. 
The difference between both protocols is stark; for $(\lambda,q)=(0.98, 0.7)$ and a desired output fidelity of approximately $0.9$, the piecemaker allows for $n=30$ end users while the factory protocol allows for only $n=5$. This highlights the potential application of the piecemaker protocol for distributing entanglement in small- and large-scale networks.

In Appendix \ref{sec:further_results} we provide for completeness further numerical comparisons, where we show the expected fidelity for both protocols as a function of $p$ and $\lambda$ in Figs.~\ref{fig:fid_vs_p} and \ref{fig:fid_vs_lambda}, respectively.

We note here that the closed-form expressions found are sensitive to numerical imprecision, due to the cancellation of large terms in the sum. This can become in particular problematic for either large $n$, or when $\lambda \approx q$. To this end we provide high-precision code available at Ref.~\cite{github_repeater-network-analytics} that implements our analytics.

\subsection{Cut-off analysis}\label{sec:cut_off_analysis}
One way to mitigate the effects of memory decoherence is by introducing cut-offs. 
A cut-off imposes constraints on how long states should be stored before resetting entanglement~\cite{rozpedek_parameter_2018, rozpedek_near-term_2019, azuma_tools_2021, goodenough_noise_2024, li2021efficient, praxmeyer_reposition_2013, shchukin_waiting_2019}. 
Note that this comes at a trade-off; when a cut-off condition is reached, entanglement is reset, leading to a reduction in the rate. 
As such, one needs to optimize over the right cut-off conditions, such as to maximize quantities that depend on both the quality and the rate.

In this section, we are concerned with a \emph{global} cut-off condition, which is governed by a parameter $T_c$. 
The parameter $T_c$ indicates the total time for which users can generate entanglement. We provide closed-form expressions for the average noise parameters subject to a cut-off $T_c$ in Appendices~\ref{sec:factory} and \ref{app:piecemaker_deriv}, see Eqs.~\eqref{eq:factory_avg_cutoff} and~\eqref{eq:piece_avg_cutoff}. Furthermore, in Appendix~\ref{app:waiting_times} we provide a simpler expression than the one found in~\cite{goodenough_noise_2024} for the average waiting time when subject to a cut-off $T_c$, see Eq.~\eqref{eq:total_waiting_time}. We note that other types of cut-offs have also been studied~\cite{praxmeyer_reposition_2013, goodenough_noise_2024}, but are outside the scope of this paper.

We now show how our analytics can be used to optimize \emph{conference key agreement rates} in star networks. 
Conference key agreement can be seen as a multipartite version of quantum key distribution, requiring the distribution of GHZ states among the end users~\cite{epping2017multi, murta2020quantum}. 
The rate at which one can perform conference key agreement depends both on the quality and the rate at which one distributes the GHZ states, see e.g.~section III.A of \cite{murta2020quantum} for more information. 
Tuning the cut-off parameter $T_c$ thus allows for a trade-off between the entanglement generation rate and the quality of the GHZ state~\cite{memmen2026strategy}.

\begin{figure}[h]
    \centering
    \includegraphics[width=1\linewidth]{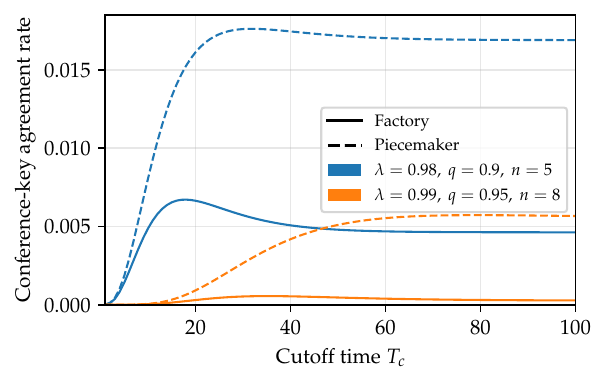}
\caption{Conference-key agreement rate as a function of the cut-off time $T_c$ for the factory and piecemaker protocol, for $\left(\lambda, q, n\right) = \left(0.98, 0.9, 5\right), \left(0.99, 0.95, 8\right)$ and dephasing noise.}
    \label{fig:cut_off}
\end{figure}

\begin{figure}[h]
    \centering
    \includegraphics[width=1\linewidth]{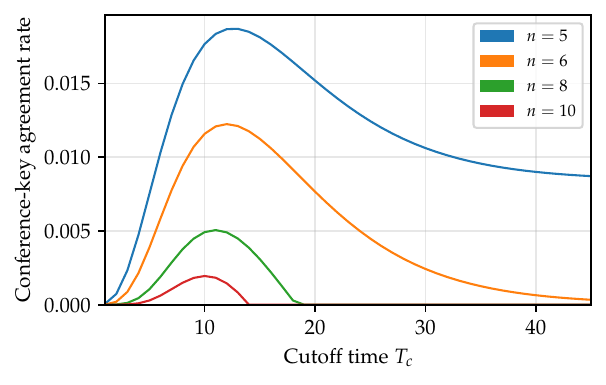}
\caption{Conference-key agreement rate as a function of the cut-off time $T_c$ for the factory protocol, for $\left(\lambda, q\right) = \left(0.99, 0.85\right)$, $n=5,6, 8, 10$ and depolarizing noise. Note that for $n\geq 6$ the use of a cut-off is necessary to generate key.}
    \label{fig:cut_off_depol}
\end{figure}

In Fig.~\ref{fig:cut_off} we show the conference key agreement rate as a function of the cut-off parameter $T_c$ for values of $\left(\lambda, q, n\right) = (0.98, 0.9, 5)$ and $(0.99, 0.95, 8)$ for the factory and piecemaker protocol. With the analytics in hand it is easy to optimize the conference-key rate by tuning the cut-off parameter $T_c$, without requiring Monte Carlo simulations. Optimizations over the cut-off $T_c$ become increasingly important for more complex noise such as depolarizing noise, since in such cases there can be a maximum storage time after which conference key cannot be generated~\cite{rozpedek_parameter_2018, memmen2026strategy}. We observe this behavior in Fig.~\ref{fig:cut_off_depol}, where we show the conference-key agreement rate as a function of the cut-off for $\left(\lambda, q\right) = \left(0.99, 0.85\right)$ and $n=5, 6, 8, 10$ for the factory protocol with depolarizing noise, where a cut-off is necessary for $n\geq 6$ to establish a non-zero conference-key agreement rate.

We note that for more complicated settings (i.e.~more complicated network topologies, inhomogeneous parameters and/or more complex distribution protocols) one can use stochastic automatic optimization techniques~\cite{avis2025optimization}.

\subsection{Distribution of the noise}\label{sec:distribution}
In the previous section, we calculated the expectation value of the noise parameter $\Lambda$. 
Although knowing the expectation value of the noise is important, the full distribution provides even more information, and is important for the design of future quantum networks and their implementations~\cite{shchukin2019waiting, li_efficient_2021, goodenough_noise_2024}. 
Since future quantum communication protocols will distribute entanglement using multiple network nodes, their design requires an understanding of the full distribution.

Rather surprisingly however, it is possible to calculate the distribution given only the expectation value $\mathds{E}\left[\Lambda\right]$ (when interpreted as a probability generating function). 
This follows from the fact that $\mathds{E}\left[\Lambda\right]=\sum_{k=0}P\left(\Lambda =\lambda^k\right)\lambda^{k}$, such that taking the $k$'th derivative, setting $\lambda$ equal to zero, and dividing by $k!$ yields
\begin{align}
\textrm{Pr}\left[\Lambda =\lambda^k\right] =&~\frac{ \left.\left(\mathds{E}\left[\Lambda\right]^{(k)}\right)\right|_{\lambda=0}}{k!} \ .
\end{align}
Here $\mathds{E}\left[\Lambda\right]^{(k)}$ is the $k$'th derivative of $\mathds{E}\left[\Lambda\right]$ with respect to $\lambda$, see e.g.~\cite{goodenough_noise_2024} for a similar approach in the context of noise in quantum networks.

\begin{figure}[h]
    \centering
    \includegraphics[width=1\linewidth]{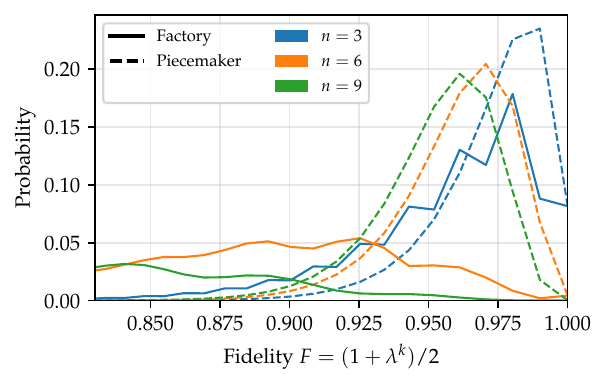}
\caption{Distribution of the fidelity for the factory and piecemaker protocol. Parameters chosen are $\lambda=0.98$, $q = 0.6$ and $n=3, 6, 9$, corresponding to the solid, dashed and dotted lines, respectively.}
    \label{fig:distribution_comparison}
\end{figure}

For small $n$, the above approach allows us to calculate the distribution analytically, while for larger $n$ the derivatives need to be calculated numerically. 
We show in Fig.~\ref{fig:distribution_comparison} the distribution for both the factory and piecemaker protocol, for values $n=3,~6,~9$ and parameters $\lambda=0.98$, $q=0.6$. 
Note also the `jaggedness' of the distributions---this is not an artifact of the numerics, but a property of the underlying distribution (see Ref.~\cite{goodenough_noise_2024} for a similar phenomenon).

The fact that the noise parameter has an underlying distribution allows us to use a so-called `binning' approach~\cite{jing2020quantum, goodenough2024near, goodenough_noise_2024}. The idea is as follows: with some probability $\textrm{Pr}\left[\Lambda =\lambda^k\right]\equiv P_k$ a state with noise parameter $\lambda^k$ is distributed. Note that after the state has been distributed, the end users can be notified which noise parameter their already measured state $\lambda^k$ has. During classical post-processing, it is then possible to group together all measurement outcomes corresponding to the same $\lambda^k$, and process them individually. A larger conference-key rate can then be achieved, since the conference-key rate is convex in $\Lambda$, i.e.~
\begin{align}
C\left(\mathds{E}\left[\Lambda\right]\right)\leq \sum_{k=0}^{\infty}P_k\cdot  C\left(\lambda^k\right) \ ,
\end{align}
where $C(\Lambda)$ is the conference-key yield for a given noise parameter $\Lambda$.
The above sum can be lower- and upperbounded by $\sum_{k=0}^{k^*}P_k\cdot  C\left(\lambda^k\right)$ and $\left(1-\sum_{k=0}^{k^*}P_k\right)+\sum_{k=0}^{k^*}P_k\cdot  C\left(\lambda^k\right)$, respectively.

\begin{figure}[h]
    \centering
    \includegraphics[width=1\linewidth]{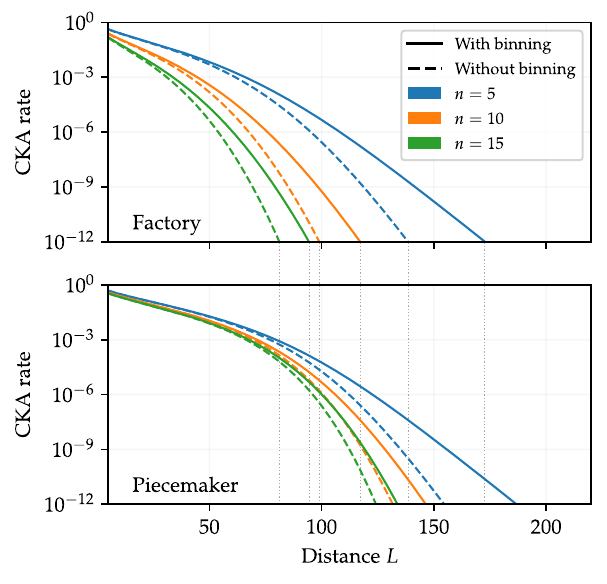}
\caption{Conference-key agreement rate with the factory protocol (top) and piecemaker protocol (bottom) as a function of the distance $L$, $\lambda=0.98$ and for $n=5, 10, 15$ end users. 
Solid lines are the case with binning (explained in the main text), dashed lines are without binning. 
The dotted, grey, vertical lines show the difference between factory and the piecemaker protocol.}
    \label{fig:binning_FP_PM}
\end{figure}

We compare the conference-key agreement rate with and without binning in Fig.~\ref{fig:binning_FP_PM} for the factory and piecemaker protocol, respectively. 
We fix $\lambda = 0.98$ and vary the distance $L$ (which determines $p$ through $\exp(-L/L_0)$ with $L_0=22$ kilometers) and number of end users $n=5,10,15$. 
We observe that the binning naturally increases the conference-key agreement rate and that the piecemaker protocol achieves significantly larger distances for the same conference-key agreement rate than the factory protocol.

\section{Conclusion}
In this paper, we studied the effect of noise on the distribution of entanglement in homogeneous star networks. 
The restriction to such a highly symmetric case allowed us to find closed-form expressions for not only the average fidelity, but also the distribution of the fidelity. Our findings confirm the numerical result in~\cite{prielinger2025piecemaker} that the piecemaker protocol outperforms the factory protocol, highlighting the piecemaker's potential in both small and large scale quantum networks. 
In fact, the piecemaker protocol already outperforms the factory protocol for $n=3$ end users, with the relative boost in fidelity increasing as the number of end users increases.

As we show in the companion paper~\cite{goodenough2026optimization}, it is possible to extend our analysis to more complex network topologies, where furthermore the parameters $\lambda, q$ need not be homogeneous. 
The numerical approach also allows us to optimize repeater placement in large-scale networks.

While a number of distribution schemes are theoretically understood~\cite{goodenough_noise_2024, kamin_exact_2023, avis2023analysis, chehimi2025entanglement}, they almost always fall under a `single-shot' setting, which we expect would not be the standard setting in future quantum networks. 
It would be of interest to see whether schemes that go beyond this paradigm are similarly amenable to analytical or numerical techniques~\cite{kunzelmann2025multiplexed}.

\section{Acknowledgements}
KG thanks Luise Prielinger for help with verifying some of the results through numerics, and Guus Avis for discussions on entanglement distribution. KG acknowledges the support from the Alexander von Humboldt Foundation. 
PE acknowledges the support received from the Dutch National Growth Fund (NGF) as part of the Quantum Delta NL program. 
PE also acknowledges the support received through the NWO-Quantum Technology program (Grant No.~NGF.1623.23.006) and funding from the Carl-Zeiss-Stiftung (CZS Center QPhoton).

\bibliography{references}

\FloatBarrier
\newpage
\onecolumngrid
\appendix
\section{Factory protocol}\label{sec:factory}
\subsection{Derivation for the factory protocol under dephasing}
We prove here a closed-form expression of the average noise in the factory protocol, in the case of dephasing. 
As before, let $n$ be the number of end users in such a network, and $\lambda, q$ be the (homogeneous) parameters describing the decoherence and the failure probability, respectively. 
A possible instance of the protocol can be described by the ordered collection $\overline{t}=\left(t_1, t_2, \ldots, t_n\right)$ of times when the $n$ links succeeded. 
To find an analytical form for the average noise parameter $\overline{\lambda}$, we start from Eq.~\eqref{eq:factory_eq} in the main text,

\begin{align}
\mathds{E}\left[\Lambda\right] =& \sum_{\overline{t}} \textrm{Pr}\left(~\overline{t}~\right)\lambda\left(\overline{t}\right)
=\left(\frac{1-q}{q}\right)^n\sum_{\overline{t}} q^{\sum_{i=1}^nt_i} \lambda^{\sum_{t_i\in\overline{t}}\left(\max(\overline{t})-t_i\right)}\ . \nonumber 
\end{align}

Ignoring the $\left(\frac{1-q}{q}\right)^n$ factor, we can rewrite the above sum as

\begin{align}
\sum_{\overline{t}} q^{\sum_{i=1}^nt_i} \lambda^{\sum_{t_i\in\overline{t}}\left(\max(\overline{t})-t_i\right)}&=
\sum_{t_\textrm{max}=1}^{\infty} \lambda^{n\cdot t_\textrm{max}} \sum_{\mathclap{\substack{\overline{t} ~\textrm{s.t.~}\\\max(\overline{t})=t_\textrm{max}}}}\left(\frac{q}{\lambda}\right)^{\sum_{i=1}^{n}t_i}\label{eq:init_expr}\\
=&\sum_{t_\textrm{max}=1}^{\infty} \lambda^{n\cdot t_\textrm{max}} \left(\left(~~~~~~\sum_{\mathclap{\substack{\overline{t} ~\textrm{s.t.~}\\\max(\overline{t})\leq t_\textrm{max}}}}\left(\frac{q}{\lambda}\right)^{\sum_{i=1}^{n}t_i}\right)\nonumber-\left(~~~~~~~\sum_{\mathclap{\substack{\overline{t} ~\textrm{s.t.~}\\\max(\overline{t})\leq t_\textrm{max}-1}}}\left(\frac{q}{\lambda}\right)^{\sum_{i=1}^{n}t_i}\right)\right)\nonumber\\
=&\left(\frac{\frac{q}{\lambda}}{1-\frac{q}{\lambda}}\right)^n\sum_{t_\textrm{max}=1}^{\infty} \lambda^{n\cdot t_\textrm{max}}\left(\left(1-\left(\frac{q}{\lambda}\right)^{t_\textrm{max}}\right)^n\nonumber-\left(1-\left(\frac{q}{\lambda}\right)^{t_\textrm{max}-1}\right)^n\right) \nonumber ,
\end{align}
where in the second equality we used that $\lbrace{\overline{t}\mid\max(\overline{t})=t_\textrm{max}\rbrace}=\lbrace{\overline{t}\mid\max(\overline{t})\leq t_\textrm{max}\rbrace}-\lbrace{\overline{t}\mid\max(\overline{t})\leq t_\textrm{max}-1\rbrace}$ and we used the geometric series in the third equality.\\

Let us drop the prefactor of $\left(\frac{\frac{q}{\lambda}}{1-\frac{q}{\lambda}}\right)^n$. Using the binomial theorem we find that

\begin{gather}
\left(1-\left(\frac{q}{\lambda}\right)^{t_\textrm{max}}\right)^n= \sum_{k=0}^{n}\binom{n}{k}(-1)^k\left(\left(\frac{q}{\lambda}\right)^{t_\textrm{max}}\right)^k\nonumber=\sum_{k=0}^{n}\binom{n}{k}(-1)^k\left(\left(\frac{q}{\lambda}\right)^{k}\right)^{t_\textrm{max}}\ ,\label{eq:binom_rewrite}\textrm{ and}\\
\left(1-\left(\frac{q}{\lambda}\right)^{t_\textrm{max}-1}\right)^n= \sum_{k=0}^{n}\binom{n}{k}(-1)^k\left(\left(\frac{q}{\lambda}\right)^{k}\right)^{t_\textrm{max}-1}\ \nonumber \ . 
\end{gather}
Combining the above two results yields
\begin{align}
    \left(1-\left(\frac{q}{\lambda}\right)^{t_\textrm{max}}\right)^n-\left(1-\left(\frac{q}{\lambda}\right)^{t_\textrm{max}-1}\right)^n\nonumber    =&~\sum_{k=0}^{n}\binom{n}{k}(-1)^k\left(\left(\left(\frac{q}{\lambda}\right)^{k}\right)^{t_\textrm{max}}-\left(\left(\frac{q}{\lambda}\right)^{k}\right)^{t_\textrm{max}-1}\right)\nonumber\\
    =&\sum_{k=0}^{n}\binom{n}{k}(-1)^k\left(\left(\frac{q}{\lambda}\right)^{k}\right)^{t_\textrm{max}}\left(1-\left(\frac{\lambda}{q}\right)^k\right)\ \nonumber,
\end{align}

Let us set the shorthand $D(k) \equiv \left(1-\left(\frac{\lambda}{q}\right)^k\right)$. We find that

\begin{align}
\sum_{t_\textrm{max}=1}^{\infty} \lambda^{n\cdot t_\textrm{max}}\left(\left(1-\left(\frac{q}{\lambda}\right)^{t_\textrm{max}}\right)^n-\left(1-\left(\frac{q}{\lambda}\right)^{t_\textrm{max}-1}\right)^n\right) \nonumber
=& \sum_{t_\textrm{max}=1}^{\infty}\sum_{k=0}^{n} \lambda^{n\cdot t_{\textrm{max}}}\, D(k)\binom{n}{k} (-1)^k \left(\left(\frac{q}{\lambda}\right)^k\right)^{t_\textrm{max}}\nonumber \\
=& \sum_{k=0}^{n} D(k)\binom{n}{k} (-1)^k \sum_{t_\textrm{max}=1}^{\infty} \left(\lambda^{n}\right)^{ t_\textrm{max}} \left(\left(\frac{q}{\lambda}\right)^k\right)^{t_\textrm{max}}\nonumber \\
=& \sum_{k=0}^{n}D(k)\binom{n}{k} (-1)^k \sum_{t_\textrm{max}=1}^{\infty} \left(\lambda^{n-k}q^k\right)^{t_\textrm{max}}\label{eq:infinite_sum1} \\
=& \sum_{k=0}^{n}D(k)\binom{n}{k} (-1)^k \left(\frac{\lambda^{n-k}q^k}{1-\lambda^{n-k}q^k}\right)\ \nonumber .\
\end{align}

Reintroducing the dropped multiplicative factors of $\left(\frac{1-q}{q}\right)^n$ and $\left(\frac{\frac{q}{\lambda}}{1-\frac{q}{\lambda}}\right)^n$ yields

\begin{align}
\mathds{E}\left[\Lambda\right] =& \left(\frac{1-q}{q}\right)^n\left(\frac{\frac{q}{\lambda}}{1-\frac{q}{\lambda}}\right)^n\sum_{k=0}^{n}D(k)\binom{n}{k} (-1)^k \left(\frac{\lambda^{n-k}q^k}{1-\lambda^{n-k}q^k}\right)\ \nonumber \nonumber\\
=&\left(\frac{1-q}{\lambda-q}\right)^n\sum_{k=0}^{n}\binom{n}{k} (-1)^k \left(1-\left(\frac{\lambda}{q}\right)^k\right)\left(\frac{\lambda^{n-k}q^k}{1-\lambda^{n-k}q^k}\right)\ \label{eq:factory_avg}.
\end{align}
Furthermore, by restricting the sum in Eq.~\eqref{eq:infinite_sum1} to a finite global cut-off $T_c$, we find that 

\begin{align}
\left(1-q^{T_c}\right)^n\mathds{E}\left[\Lambda\right]
=&\left(\frac{1-q}{\lambda-q}\right)^n\sum_{k=0}^{n}\binom{n}{k} (-1)^k \left(1-\left(\frac{\lambda}{q}\right)^k\right)\left(\frac{\lambda^{n-k}q^k}{1-\lambda^{n-k}q^k}\right)\left(1-\left(\lambda^{n-k}q^k\right)^{T_c}\right)\ \label{eq:factory_avg_cutoff},
\end{align}
since $\sum_{t_\textrm{max}=1}^{T_c}\left(\lambda^{n-k}q^k\right)^{t_\textrm{max}}= \left(\frac{\lambda^{n-k}q^k}{1-\lambda^{n-k}q^k}\right)\left(1-\left(\lambda^{n-k}q^k\right)^{T_c}\right)$, and where the factor of $\left(1-q^{T_c}\right)^n$ corresponds to conditioning on the fact that all links succeed before $T_c$. Note that we moved the term $\left(1-q^{T_c}\right)^n$ to the left-hand side to keep the expression compact.

We now deal with the limit of $q\rightarrow\lambda$ for completeness. Let us shorten $t_\textrm{max}=m$ and 

\begin{align}
\lim_{q\rightarrow\lambda}\sum_{\overline{t}} q^{\sum_{i=1}^nt_i} \lambda^{\sum_{t_i\in\overline{t}}\left(\max(\overline{t})-t_i\right)}\nonumber
=&\sum_{m=1}^\infty\lambda^{n\cdot m}\sum_{\mathclap{\substack{\overline{t} ~\textrm{s.t.~}\\\max(\overline{t})=m}}} 1 \nonumber\\
=&\sum_{m=1}^\infty\lambda^{n\cdot m}\left(m^n-\left(m-1\right)^n\right) \nonumber \\
=&\left(\sum_{m=1}^\infty\lambda^{n\cdot m}m^n\right)-\lambda^n\left(\sum_{m=0}^\infty\lambda^{n\cdot m}m^n\right) \nonumber \\
=&\left(1-\lambda^n\right)\textrm{Li}_{-n}\left(\lambda^n\right)\ \nonumber  ,
\end{align}
where $\textrm{Li}_{s}\left(z\right)\equiv \sum_{m=1}^\infty \frac{z^m}{m^s}$ is the polylogarithm of order $s$ and argument $z$. For negative integer order the polylogarithm can be expressed in terms of Eulerian numbers $\genfrac{\langle}{\rangle}{0pt}{}{n}{k}$ (see e.g.~proposition 1.4.4 of~\cite{stanley2011enumerative}), such that

\begin{align}
\mathds{E}\left[\Lambda\right] &= \left(\frac{1-\lambda}{\lambda}\right)^n\left(1-\lambda^n\right)\cdot \textrm{Li}_{-n}\left(\lambda^n\right) \nonumber \\
&=\left(\frac{1-\lambda}{\lambda}\right)^n\left(1-\lambda^n\right)\cdot \frac{1}{\left(1-\lambda^n\right)^{n+1}}\sum_{k=0}^{n}\genfrac{\langle}{\rangle}{0pt}{}{n}{k}\left(\lambda^n\right)^{n-k} \nonumber \\
&=\frac{1}{\left(1-\lambda^n\right)^n}\left(\frac{1-\lambda}{\lambda}\right)^n\cdot\sum_{k=0}^{n}\genfrac{\langle}{\rangle}{0pt}{}{n}{k}\left(\lambda^n\right)^{n-k} \nonumber \ .
\end{align}

\subsection{Depolarizing noise for the factory protocol for arbitrary states}\label{sec:dep_factory}

We now extend the analysis from the previous subsection to uniform depolarizing on arbitrary states.
We use the fact that any $n$-qubit state can be expanded in the Pauli basis,
\begin{align}
\rho = \frac{1}{2^n}\sum_{P}\mathrm{Tr}(P\rho)\cdot P , \label{eq:pure_state}
\end{align}
where the $P$ are all $4^n$ (phaseless) Pauli strings of length $n$. Since quantum channels are linear, it suffices to understand how the noise acts on Pauli strings $P$. Depolarizing noise has a particular simple effect on Pauli strings, since for a single qubit we have that
\begin{align}
I \mapsto I, \quad X\mapsto \lambda X, \nonumber \\
Y \mapsto \lambda Y,\quad Z\mapsto \lambda Z \nonumber , 
\end{align}
under the map $\rho \mapsto \lambda \rho + \left(1-\lambda\right)\mathrm{Tr}\left(\rho\right)\frac{\mathds{I}}{2}$. Extending to the $n$-qubit case, we find that for uniform depolarizing noise Pauli strings get transformed as $P\mapsto \lambda^{w(P)} P$. Here $w(P)$ is the weight of $P$, i.e.~the number of non-identity entries in $P$.

From linearity of expectation values and the fact that we consider the homogeneous setting (both in decoherence and link success probability), we are free to consider the case where a Pauli string $P$ has support on exactly the first $m$ entries. 
This implies that a closed-form expression of the following sum is desired,

\begin{align}
\mathds{E}\left[\Lambda_{m}\right]\equiv \left(\frac{1-q}{q}\right)^n\sum_{\overline{t}} q^{\sum_{i=1}^nt_i} \lambda^{\sum_{t_i\in\overline{t}[m]}\left(\max(\overline{t})-t_i\right)}\ \nonumber .
\end{align}
Here $\overline{t}[m]\equiv \left(t_1, t_2, \ldots, t_m\right)$ are the first $m$ entries of $\overline{t}$, and the above sum should be contrasted with the first expression in Eq.~\eqref{eq:init_expr}. We then find that the average state at the end of the protocol is given by 

\begin{align}
\rho' = \frac{1}{2^n}\sum_{P}\left[\mathds{E}[\Lambda_{w(P)}]\,\mathrm{Tr}(P\rho)\right] \cdot P \ \label{eq:final_state},
\end{align}
where $w(P)$ is the number of non-identity Pauli's in $P$. Assuming that the input state is pure, we find that the average output fidelity is given by
\begin{align}
F(\rho, \rho') =& \mathrm{Tr}\left(\rho\, \rho'\right)\nonumber \\
=&\frac{1}{4^n}\mathrm{Tr}\left(\sum_{P} \mathrm{Tr}(P\rho) \cdot P \times  \sum_{P'} \mathds{E}[\Lambda_{w(P')}]\cdot \mathrm{Tr}(P'\rho) \cdot P'\right)\nonumber  \\
=&\frac{1}{2^n}\sum_{P}\mathrm{Tr}(P\rho)^2\, \mathds{E}[\Lambda_{w(P)}]\nonumber \\
=&\frac{1}{2^n}\sum_{m=0}^{n}A_m \cdot  \mathds{E}[\Lambda_m] \label{eq:sector_length}\ ,
\end{align}
where we used the orthogonality of Pauli strings. In the last step we used the \emph{sector length} of the state $\rho$~\cite{wyderka2020characterizing, miller2023shor, vallee2026sector}, which is defined as

\begin{align}
    A_m \equiv \sum_{\mathclap{\substack{P \textrm{ s.t.~}\\ w(P) = m}}} \mathrm{Tr}\left(P\rho\right)^2 \ ,
\end{align}
and has been used in a variety of instances in quantum information theory~\cite{rains2002quantum, wyderka2020characterizing, rains1999quantum, miller2023shor, miller2024experimental, miller2026detecting, vallee2026sector}.

Let us now find a closed-form expression of $\mathds{E}\left[\Lambda_{m}\right]$. Proceeding analogously as in the case of Eq.~\eqref{eq:init_expr}, we find

\begin{align}
&\sum_{\overline{t}} q^{\sum_{i=1}^nt_i} \lambda^{\sum_{t_i\in\overline{t}[m]}\left(\max(\overline{t})-t_i\right)}\nonumber\\
=&\sum_{t_{\textrm{max}}=1}^\infty {\lambda^{m\cdot t_{\textrm{max}}}}\,\sum_{\mathclap{\substack{\overline{t}~\textrm{s.t.~}\\\max{\overline{t}
=t_{\textrm{max}
}}}}}\left(\frac{q}{\lambda}\right)^{\sum_{i=1}^mt_i}\,\cdot\, q^{\sum_{i=m+1}^nt_i}\nonumber \\
= &\sum_{t_{\textrm{max}}=1}^\infty \lambda^{m\cdot t_{\textrm{max}}} \, \left(\left(~~~~~~\sum_{\mathclap{\substack{\overline{t} ~\textrm{s.t.~}\\\max(\overline{t})\leq t_\textrm{max}}}}\left(\frac{q}{\lambda}\right)^{\sum_{i=1}^{m}t_i}\cdot q^{\sum_{i=m+1}^nt_i}\right)-\left(~~~~~~~\sum_{\mathclap{\substack{\overline{t} ~\textrm{s.t.~}\\\max(\overline{t})\leq t_\textrm{max}-1}}}\left(\frac{q}{\lambda}\right)^{\sum_{i=1}^{m}t_i}\cdot q^{\sum_{i=m+1}^nt_i}\right)\right)\nonumber\\
=&
\left(\frac{\frac q\lambda}{1-\frac q\lambda}\right)^m
\left(\frac{q}{1-q}\right)^{n-m}
\sum_{t_{\max}=1}^{\infty}
\lambda^{m\cdot  t_{\max}}
\left[
\left(1-\left(\frac q\lambda\right)^{t_{\max}}\right)^m
\left(1-q^{t_{\max}}\right)^{n-m}
-\left(1-\left(\frac q\lambda\right)^{t_{\max}-1}\right)^m
\left(1-q^{t_{\max}-1}\right)^{n-m}
\right] ,\label{eq:expr1}
\end{align}
where we used once more that $\lbrace{\overline{t}\mid\max(\overline{t})=t_\textrm{max}\rbrace}=\lbrace{\overline{t}\mid\max(\overline{t})\leq t_\textrm{max}\rbrace}-\lbrace{\overline{t}\mid\max(\overline{t})\leq t_\textrm{max}-1\rbrace}$, see Appendix~\ref{sec:factory}.

Using the same reasoning as in Eq.~\eqref{eq:binom_rewrite} we find that

\begin{align}
\left(1-\left(\frac q\lambda\right)^{t_{\max}}\right)^m
\left(1-q^{t_{\max}}\right)^{n-m} = \sum_{a=0}^m\sum_{b=0}^{n-m}\binom{m}{a}\binom{n-m}{b}(-1)^{a+b} \left(\frac{q^{a+b}}{\lambda^a}\right)^{t_{\max}}\nonumber \\
\left(1-\left(\frac q\lambda\right)^{t_{\max}-1}\right)^m
\left(1-q^{t_{\max}-1}\right)^{n-m} = \sum_{a=0}^m\sum_{b=0}^{n-m}\binom{m}{a}\binom{n-m}{b}(-1)^{a+b} \left(\frac{q^{a+b}}{\lambda^a}\right)^{t_{\max}-1}\ . \nonumber
\end{align}
With the above two expressions we find that
\begin{align}
&
\left(1-\left(\frac q\lambda\right)^{t_{\max}}\right)^m
\left(1-q^{t_{\max}}\right)^{n-m}
-\left(1-\left(\frac q\lambda\right)^{t_{\max}-1}\right)^m
\left(1-q^{t_{\max}-1}\right)^{n-m}\nonumber\\
=&\sum_{a=0}^m\sum_{b=0}^{n-m}\binom{m}{a}\binom{n-m}{b}(-1)^{a+b} \left(\frac{q^{a+b}}{\lambda^a}\right)^{t_{\max}}\left(1-\frac{\lambda^{a}}{q^{a+b}}\right) \ . \nonumber
\end{align}

Inserting this back into Eq.~\eqref{eq:expr1} (and momentarily dropping the prefactor of $\left(\frac{\frac q\lambda}{1-\frac q\lambda}\right)^m
\left(\frac{q}{1-q}\right)^{n-m}$) yields

\begin{align}
&\sum_{t_{\max}=1}^{\infty}
\lambda^{m\cdot  t_{\max}}
\left[
\left(1-\left(\frac q\lambda\right)^{t_{\max}}\right)^m
\left(1-q^{t_{\max}}\right)^{n-m}
-\left(1-\left(\frac q\lambda\right)^{t_{\max}-1}\right)^m
\left(1-q^{t_{\max}-1}\right)^{n-m}
\right]\nonumber \\
=&\sum_{t_{\max}=1}^{\infty}
\lambda^{m\cdot  t_{\max}} \left[\sum_{a=0}^m\sum_{b=0}^{n-m}\binom{m}{a}\binom{n-m}{b}(-1)^{a+b} \left(\frac{q^{a+b}}{\lambda^a}\right)^{t_{\max}}\left(1-\frac{\lambda^{a}}{q^{a+b}}\right)\right]\nonumber \\
=& \sum_{a=0}^m\sum_{b=0}^{n-m}\binom{m}{a}\binom{n-m}{b}(-1)^{a+b} \left(1-\frac{\lambda^{a}}{q^{a+b}}\right) \sum_{t_{\max}=1}^{\infty}
\left( \lambda^{m-a} q^{a+b}\right)^{t_{\max}} \label{eq:deriv_factory123} \\
=& \sum_{a=0}^m\sum_{b=0}^{n-m}\binom{m}{a}\binom{n-m}{b}(-1)^{a+b} \left(1-\frac{\lambda^{a}}{q^{a+b}}\right) \left(\frac{\lambda^{m-a}q^{a+b}}{1-\lambda^{m-a}q^{a+b}}\right) \ \nonumber  .
\end{align}

Reintroducing the two prefactors of $\left(\frac{1-q}{q}\right)^n \times  \left(\frac{\frac q\lambda}{1-\frac q\lambda}\right)^m
\left(\frac{q}{1-q}\right)^{n-m} = \left(\frac{1-q}{\lambda-q}\right)^m$ yields then that
\begin{align}
\mathds{E}\left[\Lambda_{m}\right] = \left(\frac{1-q}{\lambda-q}\right)^m
\sum_{a=0}^m\sum_{b=0}^{n-m}
\binom{m}{a}\binom{n-m}{b}(-1)^{a+b}
\left(1-\frac{\lambda^a}{q^{a+b}}\right)
\left(\frac{\lambda^{m-a}q^{a+b}}{1-\lambda^{m-a}q^{a+b}}\right). \label{eq:factory_avg_dep}
\end{align}

It is now possible to recover the average state $\rho'$ and average fidelity using Eqs.~\eqref{eq:final_state} and~\eqref{eq:sector_length}, respectively.

By considering a finite cut-off $T_c$ in Eq.~\eqref{eq:deriv_factory123}, we find that the expectation value for the case of a cut-off is given by

\begin{align}
&\left(1-q^{T_c}\right)^n\mathds{E}\left[\Lambda_{m}\right]\nonumber \\
= &\left(\frac{1-q}{\lambda-q}\right)^m
\sum_{a=0}^m\sum_{b=0}^{n-m}
\binom{m}{a}\binom{n-m}{b}(-1)^{a+b}
\left(1-\frac{\lambda^a}{q^{a+b}}\right)
\left(\frac{\lambda^{m-a}q^{a+b}}{1-\lambda^{m-a}q^{a+b}}\left(1-\left(\lambda^{m-a}q^{a+b}\right)^{T_c}\right)\right). \label{eq:factory_avg_dep_cutoff}
\end{align}

\section{Derivation for the piecemaker protocol}\label{app:piecemaker_deriv}
We derive now the average noise parameter for the piecemaker protocol. We start from Eq.~\eqref{eq:piecemaker_eq} in the main text, i.e.~
\begin{align}
\mathds{E}\left[\Lambda\right]=\sum_{a\geq b}\lambda^{a-b}~\textrm{Pr}\left[\max(\overline{t})=a\land \min(\overline{t})=b \right] \label{eq:sum}\ .
\end{align}

Let us first consider the case where $a=b$. Since $\lambda^{a-b} = 1$ and $\Pr\left[\max(\overline{t})=\min(\overline{t})=a\right]=\left(\frac{1-q}{q}\right)^nq^{n\cdot a}$, we find that the terms corresponding to $a=b$ contribute a total of $\left(\frac{1-q}{q}\right)^n\sum_{a=1}^{\infty}q^{n\cdot a}=\frac{\left(1-q\right)^n}{1-q^n}$ to Eq.~\eqref{eq:sum}. 

Now consider the cases where $a>b$. Note that

\begin{align}
\Pr\left[\max(\overline{t}) = a \land \min(\overline{t}) = b \right] \nonumber 
&=\Pr\left[\max(\overline{t}) \leq a \land \min(\overline{t}) \geq b\right] - \Pr\left[\max(\overline{t}) \leq a \land \min(\overline{t}) \geq b+1\right]\nonumber \\
&-\Pr\left[\max(\overline{t}) \leq a - 1 \land \min(\overline{t}) \geq b\right]
+ \Pr\left[\max(\overline{t}) \leq a - 1 \land \min(\overline{t}) \geq b+1\right]\nonumber \\
&=\left(q^{b-1}-q^a\right)^n-\left(q^{b}-q^a\right)^n-\left(q^{b-1}-q^{a-1}\right)^n+\left(q^{b}-q^{a-1}\right)^n\nonumber \\
&=f(a, b)-f(a-1, b) - f(a, b+1)+f(a-1, b+1)\nonumber \ \ ,
\end{align}
where we have set $f(a, b) \equiv \left(q^{b-1}-q^a\right)^n$. 
Expanding an $f(a, b)$ term yields \begin{align}
f(a,b)=&\sum_{k=0}^n\binom{n}{k}(-1)^k\left(q^{b-1}\right)^{n-k}\left(q^{a}\right)^{k}\nonumber \\
=&\sum_{k=0}^n\binom{n}{k}(-1)^k\left(q^{n-k}\right)^{b-1}\left(q^{k}\right)^{a}\nonumber \\
=&\sum_{k=0}^n\binom{n}{k}(-1)^k~g(a,b)\nonumber \ 
\end{align}
where we set $g(a,b)\equiv \left(q^{n-k}\right)^{b-1}\left(q^{k}\right)^{a}$. Consider now

\begin{align}
    \sum_{a> b}\lambda^{a-b}g(a, b) =& \sum_{a=1}^{\infty}\sum_{b=1}^{a-1}\lambda^{a-b} \left(q^{n-k}\right)^{b-1}\left(q^{k}\right)^{a}\nonumber \\
=&\sum_{a=1}^{\infty}\frac{\left(\lambda q^{k}\right)^{a}}{q^{n-k}}\sum_{b=1}^{a-1}\left(\frac{q^{n-k}}{\lambda}\right)^{b}\nonumber \\
=&\sum_{a=1}^{\infty}\frac{\left(\lambda q^{k}\right)^{a}}{q^{n-k}}\left(\frac{q^{n-k}}{\lambda}\cdot\frac{1-\left(\frac{q^{n-k}}{\lambda}\right)^{a-1}}{1-\frac{q^{n-k}}{\lambda}}\right)\nonumber \\
=&\frac{1}{\lambda-q^{n-k}}\sum_{a=1}^{\infty}\left(\lambda q^{k}\right)^{a}\left(1-\left(\frac{q^{n-k}}{\lambda}\right)^{a-1}\right)\nonumber \\
=&\frac{1}{\lambda-q^{n-k}}\sum_{a=1}^{\infty}\left[\left(\lambda q^{k}\right)^{a}-\frac{\lambda}{q^{n-k}}\left(q^n\right)^a\right]\label{eq:infinite_sum2}\\
=&\frac{1}{\lambda-q^{n-k}}\left[\frac{\lambda q^{k}}{1-\lambda q^{k}}-\frac{\lambda}{q^{n-k}}\cdot \frac{q^n}{1-q^n}\right]\nonumber \\
=&\frac{\lambda q^{2k}}{\left(1-q^n\right)\left(1-\lambda q^k\right)}\ \nonumber .
\end{align}

The above calculation corresponded to only including the $g(a,b)$ term. We can consider the other terms as well by noting that $g(a-1, b)=\frac{1}{q^k}g(a,b)$, and similar statements hold for $g(a,b+1)$ and $g(a-1, b+1)$. This yields $g(a,b)-g(a-1,b)-g(a,b+1)+g(a-1,b+1)= g(a,b)\left(1-\frac{1}{q^k}-q^{n-k}+\frac{q^{n-k}}{q^k}\right)= g(a,b)\frac{(1 - q^k) (q^n - q^k)}{q^{2k}}$.

From the above it follows that

\begin{align}
&\sum_{a>b}\lambda^{a-b}\left(f(a,b)-f(a-1,b)-f(a,b+1)+f(a-1, b+1)\right)\nonumber \\
=&\sum_{k=0}^n\binom{n}{k}(-1)^k\frac{\lambda q^{2k}}{\left(1-q^n\right)\left(1-\lambda q^k\right)}\cdot \frac{(1 - q^k) (q^n - q^k)}{q^{2k}}\nonumber \\
=&\,\frac{\lambda}{1-q^n} \sum_{k=0}^n\binom{n}{k}(-1)^k \frac{(1 - q^k) (q^n - q^k) }{1-\lambda q^k} \nonumber \ .
\end{align}

Hence, the final expectation value of $\Lambda$ is given by

\begin{align}
\frac{\left(1-q\right)^n}{1-q^n}+\frac{\lambda}{1-q^n} \sum_{k=0}^n\binom{n}{k}(-1)^k \frac{(1 - q^k) (q^n - q^k) }{1-\lambda q^k} \ \label{eq:piece_avg} .
\end{align}

As before, we can consider a finite global cut-off $T_c$ by restricting the sum in Eq.~\eqref{eq:infinite_sum2} to a finite one. Since
\begin{align}
&\frac{1}{\lambda-q^{n-k}}\sum_{a=1}^{T_c}\left(\lambda q^{k}\right)^{a}\left(1-\left(\frac{q^{n-k}}{\lambda}\right)^{a-1}\right)\nonumber \\
=&\left(\frac{\lambda q^{2k}}{\left(1-q^n\right)\left(1-\lambda q^k\right)}\right)\left(\frac{1-\frac{\lambda}{q^{n-k}}-q^{n\left(T_c-1\right)}\left(1-\lambda q^k\right)+\frac{1-q^n}{q^n}\left(\lambda q^k\right)^{T_c}}{1-\frac{\lambda}{q^{n-k}}}\right)\ \label{eq:piece_avg_cutoff}  ,
\end{align}

and $\left(\frac{1-q}{q}\right)^n\sum_{a=1}^{T_c}q^{n\cdot a}=\frac{\left(1-q\right)^n}{1-q^n}\left(1-q^{n\cdot T_c}\right)$, it follows that

\begin{align}
\left(1-q^{T_c}\right)^n\mathds{E}\left[\Lambda\right]=&
\frac{\left(1-q\right)^n}{1-q^n}\left(1-q^{n\cdot T_c}\right)\nonumber \\
+&\frac{\lambda}{1-q^n} \sum_{k=0}^n\binom{n}{k}(-1)^k \frac{(1 - q^k) (q^n - q^k) }{1-\lambda q^k}\left(\frac{1-\frac{\lambda}{q^{n-k}}-q^{n\left(T_c-1\right)}\left(1-\lambda q^k\right)+\frac{1-q^n}{q^n}\left(\lambda q^k\right)^{T_c}}{1-\frac{\lambda}{q^{n-k}}}\right)\nonumber \ .
\end{align}

\FloatBarrier

\section{Additional numerical results}\label{sec:further_results}
\begin{figure*}[h]
    \centering
    \begin{subfigure}[t]{0.49\textwidth}
        \centering
        \includegraphics[width=\linewidth]{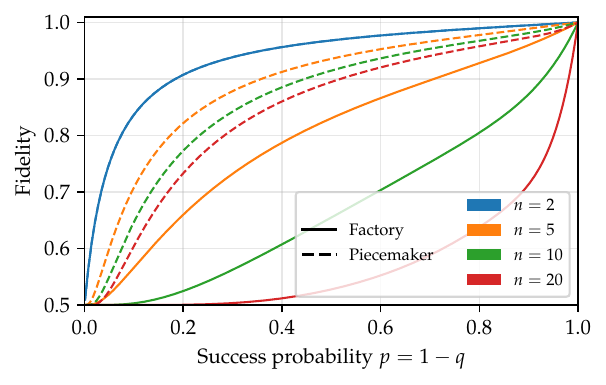}
        \caption{Fidelity as a function of the success probability $p$, for several numbers of end users $n$ and a fixed value of $\lambda = 0.95$. 
        Solid lines are for the factory protocol, while dashed lines are for the piecemaker protocol.}
        \label{fig:fid_vs_p}
    \end{subfigure}
    \hfill
    \begin{subfigure}[t]{0.49\textwidth}
        \centering
        \includegraphics[width=\linewidth]{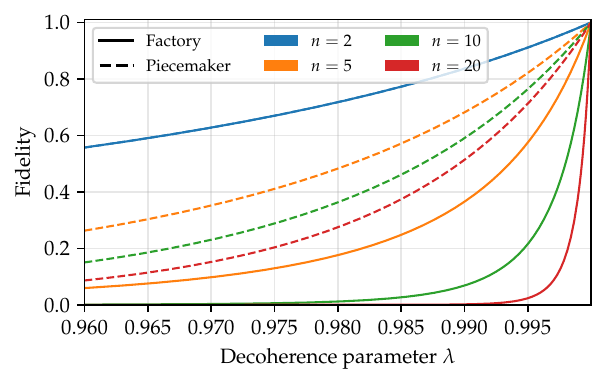}
        \caption{Fidelity as a function of the decoherence parameter $\lambda$, for several numbers of end users $n$ and a fixed value of $p= 0.95$. 
        Solid lines are for the factory protocol, while dashed lines are for the piecemaker protocol.}
        \label{fig:fid_vs_lambda}
    \end{subfigure}
    \caption{Comparison of the factory and piecemaker protocols for $n=2,5,10,20$ end users.}
    \label{fig:fidelity_comparison}
\end{figure*}

We provide here for completeness additional numerical comparisons between the factory and piecemaker protocol. We show the expected fidelity for both protocols as a function of $p$ and $\lambda$ in Figs.~\ref{fig:fid_vs_p} and \ref{fig:fid_vs_lambda}, respectively. We observe that the piecemaker protocol outperforms the factory protocol, consistent with the results from Ref.~\cite{prielinger2025piecemaker}. In particular, we confirm that the advantage of the piecemaker protocol over the factory protocol becomes more pronounced as the number of end users $n$ increases.

\FloatBarrier
\section{Waiting times in star networks with a cut-off}\label{app:waiting_times}
Here we detail the expectation value of the time until completion for the protocol, commonly called the waiting time. Note that, from the perspective of the waiting time, distributing entanglement on a star network of size $n$ takes the same amount of time as on a swap ASAP repeater chain of length $n$. It is known that for the case without a cut-off, the waiting time is given by $\sum_{k=1}^{n}\binom{n}{k}\frac{(-1)^{k+1}}{1-\left(1-p\right)^k}$~\cite{shchukin2019waiting, bernardes2011rate}.

For the case of a cut-off $T_c$ (i.e.~if not all links succeed before $T_c$, all entanglement is thrown out and the protocol starts from the beginning), the waiting time was found in~\cite{goodenough_noise_2024}, but required a sum over $T_c$ terms. We now provide a cleaner expression which does not involve such a sum over $T_c$ number of terms.


First note that the protocol consists of two parts: the failures (where not all links succeeded before $T_c$ attempts), and the success (where all links succeeded before $T_c$ attempts). The expected waiting time for the failures is given by

\begin{align}
\left(\frac{1}{\left(1-q^{T_c}\right)^n}-1\right)T_c \ ,\label{eq:waiting_time_failures}
\end{align}
so it remains to find the expected waiting time, conditioned on success. Ignoring the normalization factor of $\frac{1}{\left(1-q^{T_c}\right)^n}$ corresponding to the conditional probability for notational simplicity, we find that this waiting time is given by
\begin{align}
\left(\frac{1-q}{q}\right)^n\sum_{\overline{t}} \max(\overline{t})\cdot q^{\sum_{i=1}^nt_i} &=\left(\frac{1-q}{q}\right)^n
\sum_{t_\textrm{max}=1}^{T_c} t_\textrm{max}~\sum_{\mathclap{\substack{\overline{t} ~\textrm{s.t.~}\\\max(\overline{t})=t_\textrm{max}}}}q^{\sum_{i=1}^{n}t_i}\nonumber\\
&=\left(\frac{1-q}{q}\right)^n\sum_{t_\textrm{max}=1}^{T_c} t_\textrm{max}\left(~~~~ ~\sum_{\mathclap{\substack{\overline{t} ~\textrm{s.t.~}\\\max(\overline{t})\leq t_\textrm{max}}}}q^{\sum_{i=1}^{n}t_i}-\sum_{\mathclap{\substack{\overline{t} ~\textrm{s.t.~}\\\max(\overline{t})\leq t_\textrm{max}-1}}}q^{\sum_{i=1}^{n}t_i}\right)\nonumber\\ \nonumber\\
&=\sum_{t_\textrm{max}=1}^{T_c} t_\textrm{max}\left(\left(1-q^{t_\textrm{max}}\right)^n-\left(1-q^{t_\textrm{max}-1}\right)^n\right)\nonumber\\ 
&=\sum_{t_\textrm{max}=1}^{T_c} t_\textrm{max}\left(\left(\sum_{k=0}^{n}\binom{n}{k}(-1)^k\left(q^k\right)^{t_{\textrm{max}}}\right)-\left(\sum_{k=0}^{n}\binom{n}{k}(-1)^k\left(q^k\right)^{t_{\textrm{max}}-1}\right)\right)\nonumber\\ 
&=\sum_{t_\textrm{max}=1}^{T_c} t_\textrm{max}\left(\sum_{k=1}^{n}\binom{n}{k}(-1)^k\left(q^k\right)^{t_{\textrm{max}}}\left(1-\frac{1}{q^{k}}\right)\right)\nonumber\ . \end{align}
Exchanging sums we find the following,
\begin{align}
\sum_{k=1}^{n} \sum_{t_\textrm{max}=1}^{T_c}t_\textrm{max}\left(\binom{n}{k}(-1)^k\left(q^k\right)^{t_{\textrm{max}}}\left(1-\frac{1}{q^{k}}\right)\right)\nonumber
&=\sum_{k=1}^n \binom{n}{k}(-1)^{k+1}\frac{1-\left(1+T_c(1-q^k)\right)q^{k\cdot T_c}}{1-q^k}.
\end{align}

Reintroducing the normalization factor and adding the first term from Eq.~\eqref{eq:waiting_time_failures} we find that the total waiting time is given by

\begin{align}
&\left(\frac{1}{\left(1-q^{T_c}\right)^n}-1\right)T_c  + \frac{1}{\left(1-q^{T_c}\right)^n}\sum_{k=1}^n \binom{n}{k}(-1)^{k+1}\frac{1-\left(1+T_c(1-q^k)\right)q^{k\cdot T_c}}{1-q^k}\nonumber \\
=&\left(\frac{1}{\left(1-q^{T_c}\right)^{n}}\sum_{k=1}^{n}\binom{n}{k} (-1)^{k+1}~T_c\cdot q^{k\cdot T_c}\right) + \left(\frac{1}{\left(1-q^{T_c}\right)^n}\sum_{k=1}^n \binom{n}{k}(-1)^{k+1}\frac{1-\left(1+T_c(1-q^k)\right)q^{k\cdot T_c}}{1-q^k}\right)\nonumber \\
=&\frac{1}{\left(1-q^{T_c}\right)^n}\sum_{k=1}^n \binom{n}{k}\left(-1\right)^{k+1}\left(T_c\cdot q^{k\cdot T_c}+\frac{1-\left(1+T_c(1-q^k)\right)q^{k\cdot T_c}}{1-q^k}\right) \nonumber \\
=&~\frac{1}{\left(1-q^{T_c}\right)^n}\sum_{k=1}^n\binom{n}{k}\left(-1\right)^{k+1}\frac{1-q^{k\cdot T_c}}{1-q^k}\ \label{eq:total_waiting_time}  .
\end{align}

\end{document}